\documentclass[english,12pt]{article}
\usepackage[T1]{fontenc}
\usepackage[latin1]{inputenc}
\usepackage{a4wide}
\usepackage[pdftex]{graphicx}
\usepackage{verbatim}
\usepackage[width=0.8\textwidth]{caption}
\usepackage{amsmath,amssymb,bm}
\usepackage[english]{babel}
\usepackage[]{float}
\usepackage[]{placeins}
\usepackage{flafter}
\usepackage[]{longtable}
\usepackage{amsopn}
\usepackage{color}
\usepackage{array}
\usepackage{dsfont}
\usepackage{natbib}
\usepackage{booktabs}
\usepackage{enumerate}
\usepackage{hyperref}
\usepackage{mathtools}
\usepackage[export]{adjustbox}
\usepackage{rotating}

\makeatletter

\newtheorem{thm}{Theorem}

\newtheorem{hyp}{Assumption}

\newtheorem{rmk}{Remark}

\newcommand{\eps}{\varepsilon}

\newcommand{\indep}{\perp \!\!\! \perp}

\newcommand{\st}[1]{\texttt{#1}}

\renewcommand{\section}{\@startsection{section}{2}{0mm}{-1.5\baselineskip}{1\baselineskip}{\normalfont\large\bfseries}}
\renewcommand{\subsection}{\@startsection{subsection}{2}{0mm}{-1.2\baselineskip}{1\baselineskip}{\normalfont\normalsize\bfseries}}
\renewcommand{\subsubsection}{\@startsection{subsubsection}{3}{0mm}{-0.8\baselineskip}{0.4\baselineskip}{\normalfont\normalsize\itshape}}

\date{
\today}
\begin{document}

\title{Randomly Assigned First-Differences?\thanks{We are grateful to David Arboleda, David Autor, St\'ephane Bonhomme, Kirill Borusyak, Xavier D'Haultf\oe{}uille, Rafael Dix-Carneiro, Juan Carlos Escanciano, Yagan Hazard, Peter Hull, Thierry Mayer, Isabelle M\'ejean, Marc Melitz, and Jonathan Roth for helpful discussions. Chaisemartin was funded by the European Union (ERC, REALLYCREDIBLE, GA Number 101043899). Views and opinions expressed are those of the authors and do not reflect those of the European Union or the European Research Council Executive Agency. Neither the European Union nor the granting authority can be held responsible for them.}}

\author{Facundo Arga\~naraz\footnote{Economics Department, Sciences Po, facundo.arganaraz@sciencespo.fr.} \and Cl\'ement de Chaisemartin\footnote{Economics Department, Sciences Po, clement.dechaisemartin@sciencespo.fr.} \and Ziteng Lei\footnote{School of Labor and Human Resources, Renmin University of China, leiziteng@ruc.edu.cn.}}

\maketitle

\begin{abstract}
We consider a first-difference regression of an outcome evolution $\Delta Y$ on a treatment evolution $\Delta D$. If the treatment effect changes over time, the regression residual is a function of the period-one treatment $D_{1}$. Then, researchers should test if $\Delta D$ and $D_{1}$ are correlated: if they are, the regression may suffer from an omitted variable bias. To solve it, researchers may control for $E(\Delta D|D_{1})$. We revisit \cite{acemoglu2016import}, who study the effect of imports from China on US employment. $\Delta D$ and $D_{1}$ are correlated. $\Delta D$'s coefficient is less negative when controlling for $E(\Delta D|D_{1})$.
\end{abstract}

\newpage

\section{Introduction}

Assume that one seeks to estimate a treatment's effect with a two-period panel. If one worries that time-invariant confounders could bias a naive regression of the outcome on the treatment, one may use a first-difference (FD) regression of the outcome evolution $\Delta Y_g$ on the treatment evolution $\Delta D_g$, where $g$ indexes units of observation. Let $\widehat{\beta}$ denote the coefficient on $\Delta D_g$. The effect of the treatment may be heterogeneous, across units and/or over time. Then, a rather minimal requirement is that $\widehat{\beta}$ should be ``weakly causal'' \citep{blandhol2022tsls}: it should estimate a convex combination of unit-and-period specific effects. Under a model-based parallel-trends assumption, it is now well-known that FD regressions are typically not weakly causal \citep{dcDH2020}. One the other hand, if the treatment is binary and one is ready to assume that it is as good as random, then $\widehat{\beta}$ is weakly causal \citep{athey2021design,arkhangelsky2021double}. In this paper, we show that this design-based justification of FD regressions does not necessarily hold anymore when the treatment is non-binary and exposure to treatment varies at period one, as is often the case in practice. Also, we propose an alternative estimator that is weakly causal, under a conditional randomization assumption.

\medskip
In this introduction, let us momentarily assume that $Y_{g,t}$, the outcome of unit $g$ at $t$, is generated by the following two-way fixed effects model:
\begin{equation}\label{eq:toylinearmodel}
Y_{g,t}=\theta_g+\alpha_t+S_{t}D_{g,t}+u_{g,t},
\end{equation}
where $S_t$ is a real number. \eqref{eq:toylinearmodel} is almost the textbook linear and constant effect model \citep{angrist2009mostly,wooldridge2010econometric}, except that it allows the treatment effect $S_{t}$ to vary over time. \eqref{eq:toylinearmodel} is of course restrictive, but our main result can already be seen under that model. First-differencing \eqref{eq:toylinearmodel}, and letting $v_g=D_{g,1}(S_{2}-S_{1})+\Delta u_{g}$,
\begin{align}\label{eq:OVB_toymodel}
\Delta Y_{g}=&\theta_g+\alpha_2+S_{2}D_{g,2}+u_{g,2}-(\theta_g+\alpha_1+S_{1}D_{g,1}+u_{g,1})\nonumber\\
=&\Delta \alpha+S_{2}\Delta D_g+D_{g,1}(S_{2}-S_{1})+\Delta u_{g}\nonumber\\
=&\Delta \alpha+S_{2}\Delta D_g+v_g.
\end{align}
Then, under \eqref{eq:toylinearmodel}, the residual of a regression of $\Delta Y_{g}$ on $\Delta D_g$ is $v_g=D_{g,1}(S_{2}-S_{1})+\Delta u_{g}$, and $\Delta D_g$ has to be uncorrelated to $v_g$ for the regression not to suffer from an omitted variable bias (OVB). But if the treatment effect changes over time ($S_{2}\ne S_{1}$), $v_g$ is a function of $D_{g,1}$, so having $cov(\Delta D_g,v_g)=0$ essentially requires that $\Delta D_g$ and $D_{g,1}$ are independent. When $D_{g,1}$ is not constant, this is a strong condition: for instance, it cannot hold if $D_{g,2}$ follows a stationary AR(1) process. This is also a testable condition, and our first recommendation is that researchers invoking random assignment to justify $\widehat{\beta}$ start by regressing $\Delta D_g$ on $D_{g,1}$. This is akin to a standard balancing check. But while usually, which variables to include in balancing checks is open to interpretation, this not the case here: the residual is a function of $D_{g,1}$, so $D_{g,1}$ needs to be included.

\medskip
As the OVB is a function of the treatment-effect's evolution $S_{2}-S_{1}$, one may wonder if under reasonable assumptions this OVB simplifies, so that $\widehat{\beta}$ eventually estimates a convex combination of $S_1$ and $S_2$. We show that this is not necessarily the case: even if the treatment paths $(D_{g,1},D_{g,2})$ are randomized, which is stronger than assuming that  $\Delta D_g$ is randomized, $\widehat{\beta}$ may estimate a non-convex combination of effects. With randomly-assigned paths, estimating an FD regression is not necessary: one can merely regress $Y_{g,1}$ (resp. $Y_{g,2}$) on an intercept and $D_{g,1}$ (resp. $D_{g,2}$) to estimate the treatment effect at period one (resp. two). Thus, our result shows that under assumptions under which first-differencing is unnecessary, a FD regression may not estimate a convex combination of effects: there are instances where first-differencing can do more harm than good.

\medskip
While we consider a two-period OLS FD regression, our results can be extended to other two-way fixed effects and FD regressions.
For instance, as the coefficient on $\Delta D_{g,t}$ in a ``stacked'' FD regression with more than two time periods and period fixed effects is a weighted average of FD coefficients estimated on pairs of consecutive time periods \citep{angrist1998}, the issues we highlight also apply to stacked FD regressions.
Similarly, as the treatment coefficient in a two-way fixed effects (TWFE) regression is a weighted average of the treatment coefficients in first- and long-difference regressions \citep{ishimaru2022we}, the issues we highlight also apply to TWFE regressions.

\medskip
Empirical researchers do not seem aware that when $D_{g,1}$ varies and treatment effects vary over time, there is currently no formal result laying out assumptions under which first-difference regressions are guaranteed to be weakly causal. For example, in the survey of AER papers estimating TWFE or FD regressions conducted by \cite{dcDH2020}, three papers estimate such FD regressions where $D_{g,1}$ varies \citep{algan2010,gentzkow2011,Duranton2011}. A fourth prominent example is \cite{acemoglu2016import}, the paper we revisit. None of these papers mentions that its regression results rely on the assumption that the treatment effect is constant over time, and 3 papers estimate FD regressions separately for subsamples of time periods to investigate if effects vary over time, thus suggesting that researchers are not willing to rule out a priori such heterogeneity.\footnote{See Table 2 of \cite{acemoglu2016import}, Table 5 of \cite{gentzkow2011}, and Appendix Table 2 of \cite{Duranton2011}.}

\medskip
Moving forward, researchers using an FD regression should either acknowledge that it assumes constant effects over time and discuss the plausibility of this assumption, or they should augment their regression to ensure it remains robust to time-varying effects. For instance, we suggest regressing $\Delta Y_g$ on $\Delta D_g$, non-parametrically controlling  for $E(\Delta D_g|D_{g,1})$. If $D_{g,2}$ is randomly assigned conditional on $D_{g,1}$, the coefficient on $\Delta D_g$, $\beta_{D_1}$, identifies a variance-weighted average of conditional averages of the treatment effect at period two \citep[as in][]{angrist1998}, even if the treatment's effect changes over time. To estimate $\beta_{D_1}$, we follow \cite{chernozhukov2018double} and use a Double-Debiased Machine Learning (DDML) estimator. We also propose a placebo test of the assumption that $D_{g,2}$ is randomly assigned conditional on $D_{g,1}$. Alternatively, researchers may regress $Y_{g,2}$ on $D_{g,2}$, controlling for $E(D_{g,2}|D_{g,1},Y_{g,1})$. That regression also identifies a variance-weighted average of effects, under a sequential-randomization assumption \citep{robins1986new}.

\subsection*{Empirical example}

We apply our results to the dataset of \cite{acemoglu2016import}. We start by considering the FD regression of US industries' 1999-to-2007 employment evolutions on the evolution of their Chinese imports. The coefficient on Chinese imports is negative and highly significant, suggesting a negative effect of Chinese imports on US employment.

\medskip
First, we follow \cite{dcDH2020} and decompose the FD regression under the parallel-trends assumption. Under that assumption, $\widehat{\beta}$ estimates a weighted sum of industry$\times$year specific effects of Chinese imports on employment, where half of the effects are weighted negatively, and negative weights sum to $-0.35$. Thus, parallel-trends is not sufficient to ensure that $\widehat{\beta}$ is weakly causal. This motivates investigating whether $\widehat{\beta}$ is weakly causal under the assumption that $\Delta D$ is as good as random.

\medskip
For that purpose, we start by regressing $\Delta D_g$ on $D_{g,1}$, and we find that the two variables are strongly positively correlated: industries where imports increased the most already had larger imports in 1999. Then, $\widehat{\beta}$ could suffer from an omitted variable bias, and here is some intuition. Assume that the US-employment response to one dollar of Chinese imports becomes more negative from 1999 to 2007 ($S_2<S_1<0$ in \eqref{eq:toylinearmodel}). For instance, Chinese imports might become more substitutable to US production, as Chinese firms increase their value added \citep{chor2021growing}. Then, consider an industry $h$ with high imports $D_{h,1}$ in 1999, and an industry $\ell$ with low imports $D_{\ell,1}$. As $S_2<S_1<0$,
$D_{\ell,1}<D_{h,1}\Rightarrow D_{h,1}(S_2-S_1)<D_{\ell,1}(S_2-S_1)$. Then, even if industries' imports do not change from 1999 to 2007 (i.e. $h$ keeps importing $D_{h,1}$, and  $\ell$ keeps importing $D_{\ell,1}$), industry $h$ will experience a larger employment decrease than industry $\ell$. Therefore, due to the correlation between $\Delta D_g$ on $D_{g,1}$, the FD regression will mistakenly attribute to $\Delta D_g$ decreases in employment that would have happened without any import change, leading it to overestimate (in absolute value) the effect of those imports.

\medskip
Then, we show that even if the paths of Chinese imports $(D_{g,1},D_{g,2})$ are randomly assigned, $\widehat{\beta}$ estimates a non-convex combination of the effects of imports in 1999 and 2007, where the 1999 effect receives a weight equal to $-0.30$: assuming that Chinese imports are randomized is not enough to ensure that $\widehat{\beta}$ is weakly causal.

\medskip
Finally, we turn to our alternative estimators. Regressing $\Delta Y_g$ on $\Delta D_g$ controlling for $E(\Delta D_g|D_{g,1})$ using a DDML estimator, we obtain a significantly negative coefficient, 17 to 23\% smaller (in absolute terms) than, and significantly different from $\widehat{\beta}$. Our placebo test of the assumption that $D_{g,2}$ is randomly assigned conditional on $D_{g,1}$ is not rejected. Regressing $Y_{g,2}$ on $D_{g,2}$ controlling for $E( D_{g,2}|D_{g,1},Y_{g,1})$ yields similar results. Turning to some regressions estimated in \cite{acemoglu2016import}, controlling for $E(\Delta D_{g,t}|D_{g,t-1})$ in the stacked regression shown in the paper's Table 2 Column (2) yields a significantly negative coefficient, 33 to 38\% less negative than in the paper. Overall, while we still find evidence of a negative effect of Chinese imports on US employment, our estimated effects are meaningfully smaller than in the paper.

\subsection*{Related literature}

\cite{dcDH2020} have shown that outside of the canonical difference-in-difference design, FD regressions are not guaranteed to be weakly causal under a parallel-trends assumption. Instead, we show that FD regressions are not guaranteed to be weakly causal even if the first-differenced treatment is as good as random, a stronger assumption than parallel trends. Then, an alternative to our robust estimator $\widehat{\beta}_{D_1}$ would be to use one of the heterogeneity-robust difference-in-differences (DID) estimators that have been proposed for designs where the treatment varies at baseline \citep{dcDH2020,de2020difference,chaisemartin2022continuous}. A commonality between those DID estimators and $\widehat{\beta}_{D_1}$ is that they also control for $D_1$: they compare the outcome evolutions of units whose treatment changes (switchers) and units whose treatment does not change (stayers) with the same $D_1$. A disadvantage of the DID estimators is that they can only be used if there are stayers, a condition that does not always hold in empirical applications. An advantage of the DID estimators is that they only rely on a weaker assumption than $\widehat{\beta}_{D_1}$ . Another alternative to $\widehat{\beta}_{D_1}$ would be to use the estimator proposed in Section III of \cite{de2024difference}, that can also be used without stayers. That estimator relies on a parallel-trends assumption, and on two untestable functional-form assumptions (one on the conditional mean of the treatment effect given $D_1$ and $\Delta D$, and the other on the conditional mean of the outcome evolution without any treatment change given $D_1$). Instead, $\widehat{\beta}_{D_1}$ relies on the assumption that $D_2$ is as good as random conditional on $D_1$, and we propose a placebo test of that assumption. Moreover, $\widehat{\beta}_{D_1}$ does not rely on parametric functional-form assumptions.

\medskip
Our paper is also related to the literature considering design-based justifications of FD and TWFE regressions \citep[see e.g.][]{athey2022design,arkhangelsky2021double}. Results therein imply that with two periods and a binary randomized treatment, FD regressions always estimate a convex combination of treatment effects, even if effects vary over time. Results in \cite{borusyak2024negative} imply that this observation still holds in designs where all units are untreated at period one, and receive randomly-assigned doses of a non-binary treatment at period two. On the other hand, we show that this result no longer holds with a non-binary treatment that varies at baseline. 

\medskip
Finally, our paper is also related to the literature providing a design-based justification to shift-share (SS) IV regressions \citep{borusyak2022quasi}. Those regressions are often estimated in FD. If one directly posits a causal model in FD, the data's panel dimension can be ignored and the regression identifies a convex combination of effects \citep{borusyak2023quasi}. However, this justification only holds if there does not exist a causal model, like that in \eqref{eq:toylinearmodel}, where the outcome's level is a function of the treatment's level. If there is a causal model in levels with time-varying effects, then, as shown in \eqref{eq:OVB_toymodel}, $D_{g,1}$ is in the residual of the FD model and we are back to the issues highlighted in this paper \citep[see Theorem 3 of][who shows that the issues we highlight in this paper also apply to FD SS IV regressions]{de2021more}.\footnote{The current paper supersedes \cite{de2021more} concerning design-based justifications of FD regressions: as noted in \cite{de2021more}, results therein are not specific to FD SS IV but apply more broadly to all types of FD regressions.} Therefore, with randomly assigned FD shocks, FD SS IV regressions are only robust to time-varying effects if one rules out a causal model in levels. Researchers using that route to justify an FD SS IV regression should make this assumption explicit and explain why the theory underlying their estimation delivers a model in FD but not in levels.\footnote{See Section II.A of the
Web Appendix of \cite{autor2013china}, for an example of an economic model in levels to motivate a first-difference specification.}

\section{Set-up}\label{subsec:setup}

\paragraph{Two-period panel.} One is interested in estimating the effect of a treatment on an outcome, using a panel data set, with $G$ independent and identically distributed units indexed by $g$, and with two time periods. For $t\in \{1,2\}$, let $Y_{g,t}$ and $D_{g,t}$ respectively denote the outcome and treatment of $g$ at $t$.
\begin{hyp}
    $(Y_{g,1},Y_{g,2},D_{g,1},D_{g,2})_{g=1,...,G}$ are i.i.d.
    \label{hyp:iid}
\end{hyp}
As units are assumed i.i.d., we sometimes drop the $g$ subscript below.

\paragraph{First-difference regression.}
Let $\Delta $ denote the FD operator.
To estimate the treatment's effect, researchers often regress $\Delta Y_g$ on an intercept and $\Delta D_g$, leading to
$$\widehat{\beta}:=\frac{\sum_g (\Delta D_g-\overline{\Delta D})\Delta Y_g}{\sum_g(\Delta D_g-\overline{\Delta D})^2},$$
where $\overline{\Delta D}=1/G \sum_g \Delta D_g.$
Assuming that $(\Delta Y,\Delta D)$ has a finite second moment, under Assumption \ref{hyp:iid} the probability limit of $\widehat{\beta}$ when $G\rightarrow +\infty$ is equal to
$$\beta:=\frac{cov(\Delta D,\Delta Y)}{V(\Delta D)}.$$

\section{The FD regression may suffer from an omitted variable bias if the treatment effect changes over time.}\label{sec:results}

\paragraph{A linear causal model in levels.}
Let $Y_{g,t} (d)$  denote $g$'s potential outcome if $D_{g,t}= d$. $Y_{g,t} (0)$ is $g$'s potential outcome at $t$ if $D_{g,t}=0$, namely without treatment. Implicitly, this potential outcome notation rules out anticipation effects where $D_{g,2}$ could affect the period-one outcome, and dynamic effects where $D_{g,1}$ could affect the period-two outcome. In the introduction, we gave four examples of prominent empirical papers estimating FD regressions. Among them, none includes the lagged first-difference treatment in their regression, thus implicitly ruling out dynamic effects. That FD regressions rule out anticipation and dynamic effects is an important issue, different from that discussed in this paper. To simplify, we assume that the causal model is linear.
\begin{hyp}\label{hyp:lin_levels} For $t\in \{1,2\}$,
    	\begin{align}\label{eq:causal_model_levels}
		Y_t(d)=Y_t(0)+S_t d.
		\end{align}
\end{hyp}
Unlike in Equation \eqref{eq:toylinearmodel} in the introduction, here the slope $S_t$ is a random variable, that may vary across units. The slope is also indexed by $t$, as the treatment's effect might be time-varying.
Under \eqref{eq:causal_model_levels}, one has
\begin{equation}\label{eq:decompo_fd_under_model_levels}
\Delta Y=Y_2(0)+S_2 D_2-Y_1(0)-S_1 D_1=Y_2(0)-Y_1(0)+S_2\times \Delta D+\Delta S\times D_1,
\end{equation}
a decomposition we use repeatedly in what follows.

\paragraph{Identifying assumptions.}
In this section, we consider two identifying assumptions.
\begin{hyp}\label{hyp:random_fd}
    	\begin{align}\label{eq:indep_fd}
		\Delta D\indep (Y_2(0)-Y_1(0),S_1,S_2).
		\end{align}
\end{hyp}
Assumption \ref{hyp:random_fd} requires that the first-differenced treatment be independent of units' outcome evolutions without treatment, and of their period-one and period-two treatment effects. It is stronger than a parallel-trends assumption, that typically would only require that $Y_2(0)-Y_1(0)$ is independent or mean independent of $\Delta D$. But it is weaker than the following assumption:
\begin{hyp}\label{hyp:random_paths}
\begin{equation}\label{eq:indep_levels4}
(D_1,D_2)\indep (Y_2(0)-Y_1(0),S_1,S_2).
\end{equation}
\end{hyp}
Assumption \ref{hyp:random_paths} requires that the treatment paths $(D_1,D_2)$ be independent of units' period-one to period-two untreated potential outcome change, and of their treatment effects.

\paragraph{Decompositions of $\beta$ under Assumptions \ref{hyp:random_fd} and \ref{hyp:random_paths}.}
\begin{thm}\label{thm:OVB}
\begin{enumerate}
\item If Assumptions \ref{hyp:lin_levels} and \ref{hyp:random_fd} hold,
$$\beta=E(S_{2})+\frac{cov(\Delta D,\Delta S\times D_1)}{V(\Delta D)}.$$
\item If Assumptions \ref{hyp:lin_levels} and \ref{hyp:random_paths} hold, $$\beta=\sum_{t=1}^2\frac{V(D_t)-cov(D_1,D_2)}{\sum_{t'=1}^2[V(D_{t'})-cov(D_1,D_2)]}E(S_t).$$
\end{enumerate}
\end{thm}
\textbf{Proof}\\
\begin{align*}
\frac{cov(\Delta D,\Delta Y)}{V(\Delta D)}=&\frac{cov(\Delta D,Y_2(0)-Y_1(0))+cov(\Delta D,S_2\times \Delta D)+cov(\Delta D,\Delta S\times D_1)}{V(\Delta D)}\\
=&E(S_{2})+\frac{cov(\Delta D,\Delta S\times D_1)}{V(\Delta D)}.
\end{align*}
The first equality follows from \eqref{eq:decompo_fd_under_model_levels}. The second equality follows from Assumption \ref{hyp:random_fd}. This proves Point 1. Then,
\begin{align*}
\frac{cov(\Delta D,\Delta Y)}{V(\Delta D)}=&\frac{cov(\Delta D,Y_2(0)-Y_1(0))+cov(\Delta D,S_2\times \Delta D)+cov(\Delta D,\Delta S\times D_1)}{\sum_{t'=1}^2[V(D_{t'})-cov(D_1,D_2)]}\\
=&\frac{V(\Delta D)E(S_2)+cov(\Delta D,D_1)E(\Delta S)}{\sum_{t'=1}^2[V(D_{t'})-cov(D_1,D_2)]}\\
=&\sum_{t=1}^2\frac{V(D_t)-cov(D_1,D_2)}{\sum_{t'=1}^2[V(D_{t'})-cov(D_1,D_2)]}E(S_t).
\end{align*}
The first equality follows from \eqref{eq:decompo_fd_under_model_levels}. The second follows from Assumption \ref{hyp:random_paths}. This proves Point 2
\textbf{QED.}

\paragraph{Omitted variable bias (OVB) formula.}
Point 1 of Theorem \ref{thm:OVB} is akin to a standard omitted variable bias (OVB) formula. It shows that $\beta$ identifies the average treatment effect at period two $E(S_{2})$, plus the OVB term $$\frac{cov(\Delta D,\Delta S\times D_1)}{V(\Delta D)}.$$ If one is ready to assume that the treatment effect does not change over time ($S_2=S_1$), then the OVB is equal to zero and $\beta=E(S_{2})$. Similarly, if instead of Assumption \ref{hyp:random_fd}, one is ready to make the following stronger assumption:
\begin{align}\label{eq:indep_fd_strong}
		\Delta D\indep (Y_2(0)-Y_1(0),S_1,S_2,D_1),
		\end{align}
then the OVB is also equal to zero.
However, \eqref{eq:indep_fd_strong} implies that
\begin{align}\label{eq:indep_fd_testable}
		\Delta D\indep D_1.
		\end{align}
\eqref{eq:indep_fd_testable} is a strong condition. For instance, it cannot hold if $D_1$ and $D_2$ have the same bounded support and the distribution of $\Delta D| D_1=d$ is non-degenerate for all $d$.\footnote{Letting $\underline{d}$ and $\overline{d}$ denote the support's boundaries, the non-degenerate distribution of $\Delta D|D_1=\underline{d}$ cannot be the same as the non-degenerate distribution of $\Delta D|D_1=\overline{d}$.} If $D_{2}$ follows an AR(1) process ($D_{2}=\lambda_0+\lambda_1 D_1+\epsilon$),  \eqref{eq:indep_fd_testable} can only hold in the knife-edge case where $\lambda_1=1$, meaning that the process is non-stationary. \eqref{eq:indep_fd_testable} is also testable, for instance by regressing $\Delta D$ on $D_1$. Point 1 of Theorem \ref{thm:OVB} motivates our first recommendation: applied researchers running a first-difference regression where the baseline treatment varies should start by testing if $\Delta D$ and $D_1$ are correlated. If they are, and if the treatment effect varies over time ($S_1\ne S_2$), then the regression may be subject to an OVB.

\paragraph{Even if treatment paths are randomly assigned, $\beta$ may not estimate a convex combination of treatment effects.}
Point 2 of Theorem \ref{thm:OVB} shows that under  Assumptions \ref{hyp:lin_levels} and \ref{hyp:random_paths}, $\beta$ estimates a weighted sum of the average slopes $E(S_1)$ and $E(S_2)$. One of the two weights in Point 2 of Theorem \ref{thm:OVB} could be negative, if $V(D_t)-cov(D_1,D_2)<0$ for some $t$. The weights in Point 2 of Theorem \ref{thm:OVB} can be estimated, to assess if in a given application, $\beta$ is weakly causal or not under Assumptions \ref{hyp:lin_levels} and \ref{hyp:random_paths}.

\paragraph{First-differencing can do more harm than good.}
Point 2 of Theorem \ref{thm:OVB} still holds if Assumption \ref{hyp:random_paths} is strengthened to
\begin{equation}\label{eq:indep_levels5}
(D_1,D_2)\indep (Y_1(0),Y_2(0),S_1,S_2).
\end{equation}
Under \eqref{eq:indep_levels5}, estimating an FD regression is not necessary: one can merely regress $Y_1$ on an intercept and $D_1$ to estimate $E(S_1)$, and one can regress $Y_2$ on an intercept and $D_2$ to estimate $E(S_2)$. Thus, Point 2 of Theorem \ref{thm:OVB} shows that under assumptions under which first-differencing is unnecessary, an FD regression may not estimate a convex combination of effects. This shows that there exists instances where first-differencing can do more harm than good.

\paragraph{Special cases where the weights in Point 2 of Theorem \ref{thm:OVB} are positive.}
The weights in Point 2 of Theorem \ref{thm:OVB} are guaranteed to be positive if the treatment is binary.\footnote{$V(D_1)-cov(D_1,D_2)=E((D_1-D_2)(D_1-E(D_1)))$, and $(D_1-D_2)(D_1-E(D_1))\geq 0$ if $D_t\in \{0,1\}$.} This is consistent with the results of \cite{athey2022design} and \cite{arkhangelsky2021double}, who have shown that with a binary randomized treatment, OLS TWFE regressions always estimate a convex combination of effects. By Cauchy-Schwarz's inequality, weights are also guaranteed to be positive if $V(D_1)=V(D_2)$. Weights are also guaranteed to be positive if the treatments are not serially correlated ($cov(D_1,D_2)=0$) or negatively correlated ($cov(D_1,D_2)< 0$). Finally, weights are also guaranteed to be positive if the average treatment effect does not change over time ($E(S_1)=E(S_2)$), which is weaker than assuming that the treatment effect does not change for any unit ($S_1=S_2$). However, those positive results only hold under Assumption \ref{hyp:random_paths}, which requires that the treatment paths, rather than their first differences, be randomly assigned. Accordingly, when one of these favorable situations hold, invoking Point 2 of Theorem \ref{thm:OVB} to justify a first-difference regression requires advocating for random assignment of the paths. For instance, one should conduct balancing checks and show that pre-determined characteristics are uncorrelated with the paths.


\section{Alternative estimators}\label{sec:solution}


\subsection{Controlling for $E(\Delta D|D_1)$}

\subsubsection{Identification}

Section \ref{sec:results} shows that, under Assumption \ref{hyp:random_fd}, regressing $\Delta Y$ on $\Delta D$ may yield an OVB. This bias comes from the fact that with time-varying effects, $D_1$ is in the error term.  Notice that in the error term in \eqref{eq:decompo_fd_under_model_levels}, $D_1$ is multiplied by the random variable $\Delta S$. Then, controlling linearly for $D_1$ in the regression is generally not sufficient to solve the problem. Instead, let $\beta_{D_1}$ be the coefficient on $\Delta D$ in a regression of $\Delta Y$ on an intercept, $\Delta D$, and $E(\Delta D|D_1)$.\footnote{Of course, if $E\left(\Delta D | D_1\right)$ is linear, controlling for $E(\Delta D|D_1)$ is equivalent to controlling for $D_1$. But to avoid misspecification, we do not make this parametric functional-form assumption.} $\beta_{D_1}$ is weakly causal under the following assumption.
\begin{hyp}\label{hyp:cond_random_fd}
\begin{equation}
\Delta D \indep (Y_2(0)-Y_1(0),S_1,S_2) |D_1,
\end{equation}
\end{hyp}
which is equivalent to
\begin{equation}\label{hyp:cond_random_fda}
D_2 \indep (Y_2(0)-Y_1(0),S_1,S_2) |D_1.
\end{equation}
In \cite{acemoglu2016import}, Assumption \ref{hyp:cond_random_fd} means that among US industries with the same Chinese imports in 1999, 2007 imports are independent of $(Y_2(0)-Y_1(0))$, industries' employment evolutions without Chinese imports, and of $S_1$ and $S_2$, industries' employment elasticities with respect to Chinese imports in 1999 and in 2007. While Assumption \ref{hyp:random_fd} requires unconditional random assignment of $\Delta D$, Assumption \ref{hyp:cond_random_fd} requires random assignment conditional on $D_1$. Logically, Assumption \ref{hyp:cond_random_fd} is neither weaker nor stronger than Assumption \ref{hyp:random_fd}. In practice, conditional random assignment is usually considered as more plausible than unconditional random assignment, because the former only rules out selection on unobservables, while the latter rules out selection on observables and unobservables. Assumption \ref{hyp:random_paths} implies both Assumptions \ref{hyp:random_fd} and \ref{hyp:cond_random_fd}. On the other hand, Assumptions \ref{hyp:random_fd} and \ref{hyp:cond_random_fd} do not imply Assumption \ref{hyp:random_paths}, but situations where Assumptions \ref{hyp:random_fd} and \ref{hyp:cond_random_fd} hold while Assumption \ref{hyp:random_paths} fails seem like knife-edge cases to us. For instance, under Assumptions \ref{hyp:random_fd} and \ref{hyp:cond_random_fd},
$$0=cov(Y_2(0)-Y_1(0),\Delta D)=E(E(\Delta D|D_1)E(Y_2(0)-Y_1(0)|D_1))-E(\Delta D)E(Y_2(0)-Y_1(0)),$$
an equality that is unlikely to hold unless $D_1 \indep Y_2(0)-Y_1(0)$. But together with Assumption \ref{hyp:cond_random_fd}, $D_1 \indep Y_2(0)-Y_1(0)$ implies that $(D_1,D_2) \indep Y_2(0)-Y_1(0)$. A similar idea applies to $S_1$ and $S_2.$

\begin{thm}\label{thm:new2}
If Assumptions \ref{hyp:lin_levels} and \ref{hyp:cond_random_fd} hold,
\begin{equation}
\label{eq:betadeltad1}
\beta_{D_1}=E\left(\frac{V(D_2|D_1)}{E\left(V(D_2|D_1)\right)}E(S_2|D_1)\right).
\end{equation}
\end{thm}
\textbf{Proof}\\
Let $\Delta D_r=\Delta D-E(\Delta D|D_1)$ denote the residual from the CEF decomposition of $\Delta D$ with respect to $D_1$.
Under Assumption \ref{hyp:cond_random_fd},
\begin{align}\label{eq:thm_condrand_step1}
0&=cov(\Delta D_r,Y_2(0)-Y_1(0))=cov(\Delta D_r,S_2)=cov(\Delta D_r,S_2E(\Delta D|D_1))\nonumber\\
=&cov\left(\Delta D_r,\Delta S\times D_1\right).
\end{align}
For instance,
\begin{align*}
cov(\Delta D_r,S_2E(\Delta D|D_1))=&E(\Delta D_rS_2E(\Delta D|D_1))\\
=&E(E(\Delta D_r|S_2,D_1)S_2E(\Delta D|D_1))\\
=&E((E(\Delta D|S_2,D_1)-E(\Delta D|D_1))S_2E(\Delta D|D_1))\\
=&0.
\end{align*}
Then,
\begin{align}
\beta_{D_1}=&\frac{cov(\Delta D_r,\Delta Y)}{V(\Delta D_r)} \label{eq:betad1}\\
=&\frac{cov(\Delta D_r,Y_2(0)-Y_1(0))+cov(\Delta D_r,S_2\times \Delta D)+cov(\Delta D_r,\Delta S\times D_1)}{V(\Delta D_r)} \nonumber\\
=&\frac{cov(\Delta D_r,S_2 E(\Delta D|D_1))+cov(\Delta D_r,S_2\times \Delta D_r)}{V(\Delta D_r)}  \nonumber\\
=&\frac{E\left((\Delta D_r)^2S_2\right)}{E\left((\Delta D_r)^2\right)}  \nonumber\\
=&E\left(\frac{V(\Delta D|D_1)}{E\left(V(\Delta D|D_1)\right)}E(S_2|D_1)\right)  \nonumber \\
=& E\left(\frac{V( D_2|D_1)}{E\left(V(D_2|D_1)\right)}E(S_2|D_1)\right). \nonumber
\end{align}
The first equality follows from the Frich-Waugh-Lovell theorem. The second follows from \eqref{eq:decompo_fd_under_model_levels}. The third and fourth follow from \eqref{eq:thm_condrand_step1}. The fifth follows from the law of iterated expectations and Assumption \ref{hyp:cond_random_fd}. Finally, the last follows from the definition of $\Delta D$.


\textbf{QED.}

\medskip
Theorem \ref{thm:new2} shows that, under Assumption \ref{hyp:cond_random_fd}, $\beta_{D_1}$ identifies a weighted average of conditional average slopes $E(S_2|D_1)$. The weights are determined by the conditional variance of $\Delta D$ given $D_1$ and thus are nonnegative. When $D_1$ is discrete, those weights are the same as the conditional variance-weights that appear in the estimand from a regression of $Y_2$ on $D_2$ saturated in $D_1$ \citep[cf. Equation (3.3.7) of][]{angrist2009mostly}. Here, instead of fully-saturating the regression, we are nonparametrically controlling for $E\left(\left. D_2 \right| D_1\right),$ which is feasible even when $D_1$ is continuously distributed. Theorem \ref{thm:new_nl} in the Appendix shows that without the linearity condition in Assumption \ref{hyp:lin_levels}, if the distribution of $D_2$ conditional on $D_1$ is continuous everywhere, $\beta_{D_1}$ identifies a weighted average, across $x$ and $d_1$, of $E\left(\frac{\partial Y_2(x)}{\partial x}\middle|D_1=d_1\right)$, the conditional average of the derivative of $x\mapsto Y_2(x)$ evaluated at $x$ and among units with $D_1=d_1$. If the distribution of $D_2$ conditional on $D_1$ is discrete, one can show that $\beta_{D_1}$ identifies a weighted average of slopes. Thus, $\beta_{D_1}$ remains weakly causal even if the treatment effect is non linear.

\medskip
The following remark implies that while $\beta_{D_1}$ is weakly causal under Assumptions \ref{hyp:lin_levels} and \ref{hyp:cond_random_fd}, those assumptions are not sufficient to ensure that $\beta$ is weakly causal.
\begin{rmk}
	\label{rmk:betaunderA5}
	Under Assumptions \ref{hyp:lin_levels} and \ref{hyp:cond_random_fd}, and if $\Delta D \indep (Y_2(0) - Y_1(0)),$ then
	\begin{equation}
		\label{eq:betaunderA5}
		\beta = \sum^2_{t = 1} E\left(\omega_{t}\left(D_1\right) E\left(\left. S_t \right| D_1\right) \right),
	\end{equation}
	where
	\begin{equation*}
		\begin{split}
				\omega_1\left(D_1\right) & = \frac{D_1^{2} - E\left(D_1\right)D_1 - \left(E\left(\left. D_2 \right|D_1\right)D_1 - E\left(D_2 \right)D_1\right)}{\sum^2_{t^{\prime}=1} \left[V\left(D_{t^{\prime}}\right) - cov\left(D_1, D_2\right)\right]}, \\
					\omega_2\left(D_1\right) & = \frac{E\left(\left. D_2^2 \right| D_1\right) - E\left(D_2\right) E\left(\left. D_2 \right| D_1\right) - \left(E\left(\left. D_2 \right| D_1\right)D_1 - E\left(\left. D_2 \right| D_1\right)E\left(D_1\right)\right)}{\sum^2_{t^{\prime}=1} \left[V\left(D_{t^{\prime}}\right) - cov\left(D_1, D_2\right) \right]}.
		\end{split}
	\end{equation*}
	Hence, $\beta$ can be written as a weighted sum, across $D_1$ and $t$, of the conditional average slopes $E\left(\left. S_t \right| D_1\right)$. $E(\omega_{t}\left(D_1\right))=(V(D_t)-cov(D_1,D_2))/\sum^2_{t^{\prime}=1} \left[V\left(D_{t^{\prime}}\right) - cov\left(D_1, D_2\right)\right]$, which may be negative for $t=1$ or $t=2$. Therefore, the weights $\omega_{t}\left(D_1\right)$ are not necessarily almost-surely nonnegative. Thus, Assumptions \ref{hyp:lin_levels} and \ref{hyp:cond_random_fd} are not sufficient to ensure that $\beta$ is weakly causal, even under the additional condition that $\Delta D \indep (Y_2(0) - Y_1(0))$.\footnote{Without that additional condition, $\beta$ would also suffer from an OVB, on top of not being weakly causal.}
\end{rmk}

\textbf{Proof}\\
Recall that $\beta$ is defined as
$$
\beta = \frac{cov\left(\Delta D, \Delta Y\right)}{V\left(\Delta D\right)}.
$$
$V\left(\Delta D\right)=\sum^2_{t^{\prime}=1} \left[V\left(D_{t^{\prime}}\right) - cov\left(D_1, D_2\right) \right]$. Then, by Assumption \ref{hyp:lin_levels} and $\Delta D\indep (Y_2(0) - Y_1(0)),$
\begin{equation}
	\label{eq:numbetaA5}
	cov\left(\Delta D, \Delta Y\right) = cov\left(\Delta D, S_2 \times \Delta D\right) + cov\left(\Delta D, \Delta S \times D_1\right).
\end{equation}
 By Assumption \ref{hyp:cond_random_fd},
 \begin{align}
 	cov\left(\Delta D, S_2 \times \Delta D\right) & = E\left(\left(E\left( \left.\Delta D^2 \right| D_1\right) - E\left(\Delta D\right) E\left(\left. \Delta D \right| D_1\right) \right) E\left(\left. S_2 \right| D_1\right)\right), \label{eq:numbetaA5ft} \\
 	cov\left(\Delta D, \Delta S \times D_1\right) & = E\left(\left(E\left(\left.\Delta D \right| D_1\right) D_1 - E\left(\Delta D\right) D_1\right) E\left(\left. \Delta S \right| D_1\right)\right). \label{eq:numbetaA5st}
 \end{align}
Plugging \eqref{eq:numbetaA5ft} and \eqref{eq:numbetaA5st} into \eqref{eq:numbetaA5} yields \eqref{eq:betaunderA5} \textbf{QED.}

\subsubsection{Estimation}
\label{subsec:ecestimation}

When $D_1$ is a discrete variable, estimation of $E(\Delta D|D_1)$
and then of $\beta_{D_1}$ is straightforward. Accordingly, in this section we focus on the case where $D_1$ is continuous.

\medskip
For estimation purposes, we first note that \eqref{eq:betad1} is equivalent to
\begin{equation}
	\label{eq:betad1lr}
\beta_{D_1} = \frac{cov\left(\Delta Y_r, \Delta D_r\right)}{V\left(\Delta D_r\right)}, 	
\end{equation}
where $\Delta Y_r = \Delta Y - E\left(\left. \Delta Y \right| D_1\right)$. Equation \eqref{eq:betad1lr} is convenient since it means that $\beta_{D_1}$  is identified by a Neyman-orthogonal moment, locally robust to both $E\left(\left. \Delta D \right| D_1\right)$ and $E\left(\left. \Delta Y \right| D_1\right)$ \citep[][]{chernozhukov2018double}. This implies that we can estimate these conditional expectations through nonparametric flexible procedures (e.g., machine learning) and inference on \eqref{eq:betad1lr} will be valid under more general conditions than plug-in approaches (using, e.g., \eqref{eq:betad1}), as discussed by \cite{chernozhukov2022locally}.

\paragraph{Estimator.} To estimate $\beta_{D_1}$,  we shall use the Double-Debiased Machine Learning (DDML) procedure of \cite{chernozhukov2018double}. Note that this is equivalent to \cite{robinson1988root}'s estimator, but its implementation requires cross-fitting. To this end, we need to randomly partition the sample into $L$ folds, $I_1, \cdots, I_L$, of the same size. Let $I^c_\ell$ be the complement of $I_\ell$ in $\left\{1,2\cdots,n\right\}$. To simplify our discussion, let $\eta\left(D_1\right) = E(\Delta D|D_1)$ and $\gamma\left(D_1\right) = E(\Delta Y|D_1)$. We conduct estimation of $\beta_{D_1}$ as follows:

\medskip \noindent \textbf{Step 1:} Estimate $\eta$ and $\gamma$ using $I^c_\ell$, obtaining cross-fitted estimators $\hat{\eta}_\ell$ and $\hat{\gamma}_\ell$, for each $\ell = 1,2,\cdots,L$.

\medskip \noindent \textbf{Step 2:} Construct residuals $\widehat{\Delta D}_{r,i} = \Delta D_i - \hat{\eta}_\ell(D_{1i})$ and $\widehat{\Delta Y}_{r,i} = \Delta Y_i - \hat{\gamma}_\ell(D_{1i})$, for each $i \in I_\ell$, $\ell = 1,2,\cdots,L$.

\medskip \noindent \textbf{Step 3:} Run OLS of $\widehat{\Delta Y}_r$ on $\widehat{\Delta D}_r$. The corresponding slope is our estimator $\widehat{\beta}_{D_1}$.

\medskip The asymptotic properties of the previous estimator follows by a straightforward application of Theorems 3.1-3.2 of \cite{chernozhukov2018double}.

\subsubsection{Placebo tests}\label{subsec:alternatives}

Placebo tests are an important step in establishing the credibility of an identifying assumption \citep{imbens2001estimating,imbens2024lalonde}. However, placebo testing Assumption \ref{hyp:cond_random_fd} is not straightforward. To see this, assume one observes another period of data, period 0, and let $\Delta Y_{-1}$, $\Delta D_{-1}$, and $\Delta S_{-1}$ respectively denote the outcome, treatment, and treatment-effect evolutions from period $0$ to $1$.
Assume also that Assumption \ref{hyp:cond_random_fd} holds with all the unobservables switched back by one period:
\begin{equation*}
\Delta D \indep (Y_1(0)-Y_0(0),S_0,S_1) |D_1.
\end{equation*}
Then, one may still have that the coefficient on $\Delta D$ in the regression of $\Delta Y_{-1}$ on an intercept, $\Delta D$, and $E(\Delta D|D_1)$ could differ from zero.
\begin{align*}
cov(\Delta D_r,\Delta Y_{-1})=&cov(\Delta D_r,Y_1(0)-Y_0(0))+cov(\Delta D_r,D_1 \Delta S_{-1})+cov(\Delta D_r,\Delta D_{-1}S_0)\\
=&cov(\Delta D_r,\Delta D_{-1}S_0),
\end{align*}
and $cov(\Delta D_r,\Delta D_{-1}S_0)$ could differ from zero, in particular if $\Delta D_r$ is correlated to $\Delta D_{-1}$.

\medskip
Instead, let $\beta_{D_0,D_1}$ denote the coefficient on $\Delta D$ in a regression of $\Delta Y$ on an intercept, $\Delta D$, and $E(\Delta D|D_0,D_1)$.
Let $\Delta D_{r,2}=\Delta D-E(\Delta D|D_0,D_1)$ denote the residual from the CEF decomposition of $\Delta D$ with respect to $(D_0,D_1)$.
Let $\beta^{pl}_{D_0,D_1}$ denote the coefficient on $\Delta D$ in a regression of $\Delta Y_{-1}$ on an intercept, $\Delta D$, and $E(\Delta D|D_0,D_1)$.
\begin{hyp}\label{hyp:cond_random_fd_placebotestable}
\begin{equation}
\Delta D \indep (Y_2(0)-Y_1(0),Y_1(0)-Y_0(0),S_0,S_1,S_2)|D_0,D_1
\end{equation}
\end{hyp}
Assumption \ref{hyp:cond_random_fd_placebotestable} generalizes Assumption \ref{hyp:cond_random_fd} to settings with three dates. Theorem \ref{thm:new3} below shows that under Assumption \ref{hyp:cond_random_fd_placebotestable},  $\beta_{D_0,D_1}$ identifies a variance-weighted average of conditional slopes. Moreover, Assumption \ref{hyp:cond_random_fd_placebotestable} has the following, testable implication: $\beta^{pl}_{D_0,D_1}=0.$
\begin{thm}\label{thm:new3}
If Assumptions \ref{hyp:lin_levels} and \ref{hyp:cond_random_fd_placebotestable} hold, $$\beta_{D_0,D_1}=E\left(\frac{V(D_2|D_0,D_1)}{E\left(V(D_2|D_0,D_1)\right)}E(S_2|D_0,D_1)\right),$$ and $\beta^{pl}_{D_0,D_1}=0.$
\end{thm}
\textbf{Proof}\\
The proof of the first equality is similar to the proof of the first equality in Theorem \ref{thm:new2}, and is therefore omitted. Turning to the second equality, it is straightforward to see that
\begin{align*}
\beta^{pl}_{D_0,D_1}=&\frac{cov(\Delta D_{r,2},\Delta Y_{-1})}{V(\Delta D_{r,2})}\\
=&\frac{cov(\Delta D_{r,2},Y_1(0)-Y_0(0))+cov(\Delta D_{r,2},S_1 \Delta D_{-1})+cov(\Delta D_{r,2},\Delta S_{-1} D_0)}{V(\Delta D_{r,2})}\\
=&\frac{cov(\Delta D_{r,2},S_1D_1)-cov(\Delta D_{r,2},S_1D_0)}{V(\Delta D_{r,2})}\\
=&0.
\end{align*}
\textbf{QED.}

\subsection{Controlling for $E(D_2|D_1,Y_1)$}
\label{sec:solution2}

Instead of Assumption \ref{hyp:cond_random_fd}, researchers may prefer to make the following assumption:
\begin{hyp}\label{hyp:cond_random_fdaug}
	\begin{equation}
		 D_2 \indep (Y_2(0), S_2) |D_1, Y_1.
	\end{equation}
\end{hyp}
Assumption \ref{hyp:cond_random_fdaug} is a
sequential-randomization assumption, as in \cite{robins1986new}, which for instance holds when $D_2$ is randomly assigned conditional on $(D_1,Y_1)$. In \cite{acemoglu2016import}, this means that among US industries with the same Chinese imports and employment in 1999, 2007 imports are as good as random.
The fact that the lagged outcome is conditioned upon in Assumption \ref{hyp:cond_random_fdaug} and not in Assumption \ref{hyp:cond_random_fd} may make the former assumption more plausible.

\medskip
Let $\beta_{D_1,Y_1}$ be the coefficient on $D_2$ in a regression of $Y_2$ on an intercept, $D_2$, and $E \left(D_2|D_1,Y_1\right)$. Also, let $D^{aug}_{2r} = D_2  - E \left(D_2|D_1,Y_1\right)$. $\beta_{D_1,Y_1}$ is weakly causal under Assumption \ref{hyp:cond_random_fdaug}:
\begin{thm}\label{thm:new4}
	If Assumptions \ref{hyp:lin_levels} and \ref{hyp:cond_random_fdaug} hold,
	\begin{equation}
		\label{eq:betad2}
		\beta_{D_1,Y_1}=E\left(\frac{V( D_2|D_1,Y_1)}{E\left(V(D_2|D_1,Y_1)\right)}E(S_2|D_1, Y_1)\right).
	\end{equation}
\end{thm}
\textbf{Proof}\\
By the law of iterated expectations and Assumption \ref{hyp:cond_random_fdaug}, one can show that $cov\left(D^{aug}_{2r}, Y_2(0)\right) = 0$. Also,
\begin{align*}
	cov\left(D^{aug}_{2r}, S_2D_2\right) & = E \left(D^{aug}_{2r} S_2D_2\right) \\
	& = E \left( E\left( \left. \left(D_2 - E \left(D_2| D_1, Y_1\right)\right) D_2 \right| D_1, Y_1\right) E\left(S_2 | D_1, Y_1\right)\right) \\ & = E \left(V \left( \left. D_2 \right| D_1, Y_1\right) E\left(S_2 | D_1, Y_1\right)\right).
\end{align*}
Similarly, by the law of iterated expectations,
\begin{align*}
	V\left(D^{aug}_{2r}\right) & = E \left( \left(D_2 - E\left( \left. D_2 \right| D_1, Y_1\right)\right)^2\right) \\
	& = E \left( V \left(\left. D_2 \right| D_1, Y_1\right)\right).
\end{align*}
As a result,
\begin{align*}
	\beta_{D_1,Y_1} & = \frac{cov( D^{aug}_{2r}, Y_2)}{V(D^{aug}_{2r})} \\ & = E\left(\frac{V( D_2|D_1,Y_1)}{E\left(V(D_2|D_1,Y_1)\right)}E(S_2|D_1, Y_1)\right).
\end{align*}
\textbf{QED.}

\medskip
To estimate $\beta_{D_1,Y_1}$, one can follow the exact same steps as in Section \ref{subsec:ecestimation}, replacing $\Delta Y$ and $\Delta D$ by $Y_2$ and $D_2$, and $E(\Delta D|D_1)$ by $E(D_2|D_1,Y_1)$.

\section{Application}\label{sec:application}

\paragraph{Reanalysing a two-period OLS FD regression.}
We use the panel data set constructed by \cite{acemoglu2016import}, who estimate FD regressions of US industries' employment evolutions on the evolution of their Chinese imports penetration ratio.
The paper's Table 2 shows several FD IV regressions (Columns (5) to (8)) and one stacked FD OLS regression with two first differences and three periods (Column (2)), but it does not contain any two-period FD OLS regression. To ensure our empirical application is closely aligned with our theoretical results, we start by revisiting one two-period FD OLS regression, before considering regressions that were estimated in the paper. We consider the 1999 to 2007 FD regression of US industries employment evolutions on the evolution of their Chinese imports, the two-period FD regression whose Chinese-imports coefficient is the closest to that in the stacked FD regression in the paper's Table 2 Column (2). Like in the paper, all linear regressions below are weighted by industries' 1991 employment, and standard errors are clustered at the three-digit industry level. As shown in Table \ref{table:AADH}, $\widehat{\beta}=-0.78$ (s.e.=$0.22$), which is very close to the coefficient in the paper's Table 2 Column (2) ($-0.81$, s.e.=$0.16$). According to this regression, an increase of Chinese imports decreases US employment.

\paragraph{Decomposing $\beta$ under the parallel-trends assumption.}
First, we follow \cite{dcDH2020}, who show that under the parallel-trends assumption, an FD regression estimates a weighted sum of the treatment's effect in each $(g,t)$ cell, with weights that can be estimated. We estimate those weights, and find that under the parallel-trends assumption, $\beta$ estimates a weighted sum of 784 industry$\times$year specific effects of Chinese imports on employment, where more than a half of effects are weighted negatively, and where negative weights sum to $-0.35$. Then, the parallel-trends assumption is not sufficient to ensure that $\beta$ is weakly causal. This motivates investigating whether $\beta$ is weakly causal under the assumption that $\Delta D$ is as good as randomly assigned.

\paragraph{Testing if $\Delta D$ and $D_1$ are correlated.}
We start by testing \eqref{eq:indep_fd_testable}, by regressing $\Delta D$, the change in industries' import-penetration ratio, on $D_1$, industries' import-penetration ratio in 1999. The coefficient on $D_1$ is equal to $0.74$, and it is highly significant  (s.e.=$0.16$). Thus, the industries that experienced the largest growth of their imports-penetration ratio from 1999 to 2007 already had a larger imports-penetration ratio in 1999.

\paragraph{Estimating the weights attached to $\beta$ in Point 2 of Theorem \ref{thm:OVB}.}
Next, we estimate the weights attached to $\beta$ in Point 2 of Theorem \ref{thm:OVB}. Under Assumptions \ref{hyp:lin_levels} and \ref{hyp:random_paths}, $\beta$ estimates a weighted sum of the average effects of import penetration in 1999 and 2007, where the 2007 average effect receives a weight equal to 1.3, while the 1999 average effect receives a weight equal to -0.3. Thus, $\beta$ is not weakly causal under Assumptions \ref{hyp:lin_levels} and \ref{hyp:random_paths}.

\paragraph{Estimation of $\beta_{D_1}$.}
We implement the algorithm outlined in Section \ref{subsec:ecestimation}, with $L=5$. We first assume that $E(\Delta D|D_1)$ is linear, that is, we estimate $\Delta Y$ on an intercept, $\Delta D$, and $D_1$, and look at the estimated coefficient on $\Delta D$.\footnote{Since the regression is fully parametric, no Neyman-orthogonal correction is needed.} When we do this, we obtain $\widehat \beta_{D_1} = -0.65$, which is statistically significant at the usual levels (s.e.=0.24, t-stat=-2.65). Moreover, we find that this estimator is statistically different for $\hat \beta$ ($H = 6.99$).\footnote{$H$ refers to the Generalized Hausman-test statistic.} Then, we estimate $E(\Delta D|D_1)$ and $E(\Delta Y|D_1)$ non-parametrically using Lasso regressions on a polynomial of degree three in $D_1$, with Stata's \st{rlasso} command and the command's default choices. With a third degree polynomial, the Lasso selects only $D_1$, for both regressions, and going beyond that does not change the selection. $\widehat{\beta}_{D_1}$ is equal to $-0.60$, it is significant (s.e.=0.21, t-stat=-2.87), and it is significantly different from $\widehat{\beta}$ ($H=6.43$). Using instead a Neural Networks (NN) estimator to estimate $E(\Delta D|D_1)$ and $E(\Delta Y|D_1)$, with Stata's \st{brain} command,\footnote{For training, it initiates with 1,000 iterations using a training factor $eta=1$. Note that an advantage of basing inference on the Neyman-orthogonal moment leading to \eqref{eq:betad1lr} is that estimation is locally insensitive to tuning parameters involved in the first-stage.} leads to a point estimate of $-0.65$, which is significant (s.e.=0.21, t-stat=-3.04) and statistically different from $\widehat{\beta}$ ($H=6.91$). In all cases, our proposed estimators indicate a less negative effect of Chinese imports on US employment than the simple first-difference estimator.

\paragraph{Estimation of $\beta_{D_1,Y_1}$.}
Instead of $\beta_{D_1}$, one may prefer to estimate
$\beta_{D_1,Y_1}$, the coefficient on $D_2$ in a regression of $Y_2$ on an intercept, $D_2$, and $E \left(D_2|D_1,Y_1\right)$. Table \ref{table:AADH2} in the appendix shows that doing so leads to results similar to those shown in Table \ref{table:AADH}.

\paragraph{Placebo test.}
We first conduct a naive placebo test by regressing $\Delta Y_{-1}$, industries employment evolutions from 1991 to 1999, on an intercept and $\Delta D$.  As shown in Table \ref{table:AADH}, the coefficient on $\Delta D$ is $-0.23$ (s.e.=0.15), which is statistically insignificant. However, this naive test could be misleading as $\Delta D$ and $\Delta D_{-1}$ are very strongly positively correlated (correlation=0.49). To perform a proper placebo test, we follow Theorem \ref{thm:new3}. First, we consider the simple case where $E(\Delta D | D_0, D_1)$ is linear in $D_0$ and $D_1$. We run a regression of $\Delta Y_{-1}$ on an intercept, $D_0$, $D_1$, and $\Delta D$. The coefficient on $\Delta D$ is small and not significantly different from zero, so Assumption \ref{hyp:cond_random_fd_placebotestable} is not rejected. Next, we estimate both $E(\Delta D|D_0,D_1)$ and $E(\Delta Y_{-1}|D_0,D_1)$ using Lasso or NN, and run OLS of the residual of $\Delta Y_{-1}$ on that of $\Delta D$.\footnote{As before, for the Lasso regressions a third degree polynomial in both variables has been specified.} In both cases, the coefficient on $\Delta D$ is small and not significantly different from zero. Overall, our placebos lend support to Assumption \ref{hyp:cond_random_fd_placebotestable}.

\paragraph{Do $\beta$ and $\beta_{D_1}$ just estimate different causal effects, or should we trust more one of the two estimands?}
$\beta$ and $\beta_{D_1}$ respectively estimate $E(S_2)$ and a variance-weighted average of the slopes $S_2$ under different, non-nested assumptions (Assumption \ref{hyp:random_fd} and $S_2=S_1$ for $\beta$, and Assumption \ref{hyp:cond_random_fd} for $\beta_{D_1}$). It does not seem very likely to us that $\beta$'s and $\beta_{D_1}$'s identifying assumptions both hold and the difference between them just comes from the fact they estimate different parameters. As explained earlier, Assumptions \ref{hyp:random_fd} and \ref{hyp:cond_random_fd} jointly hold ``essentially only if'' Assumption \ref{hyp:random_paths} holds. This condition seems very strong to us: the entire path of Chinese imports is unlikely to be randomly assigned to industries. Accordingly, it seems more likely to us that the difference between $\beta$ and $\beta_{D_1}$ comes, at least in part, from a violation of one or both of the estimators' identifying assumptions. A placebo test of the assumptions underlying $\beta_{D_1}$ is not rejected, that estimator does not require assuming constant effects over time, and Assumption \ref{hyp:cond_random_fd}, while neither weaker nor stronger than Assumption \ref{hyp:random_fd}, may be more plausible. Thus, $\beta$ seems more likely to us to suffer from a violation of its identifying assumptions, be it Assumption \ref{hyp:random_fd}, which would lead to a standard OVB, or the assumption that the treatment effect is stable over time, which would lead to the type of OVB our paper discusses. Overall, we think that $\beta_{D_1}$ is a more plausible estimand of the effect of Chinese imports of US employment than $\beta$.

\paragraph{Re-analysis of the stacked regression in the paper's Table 2 Column (2).}
The paper's Table 2 Column (2) contains a stacked FD regression with three periods and two first-differences, of $\Delta Y_t$ on $\Delta D_t$ and a fixed effect (FE) for the second FD. The coefficient on $\Delta D_t$ is equal to $-0.81$ (s.e.=$0.16$). Instead, we start by assuming that $E(\Delta Y_t|D_{t-1})$ is linear, and regress $\Delta Y_t$ on $\Delta D_t$ and $D_{t-1}$. The coefficient on $\Delta D_t$ is equal to -0.60, it is significantly different from zero (s.e.=0.15), and statistically different from the paper's coefficient ($H=7.55$). Next, we regress $\Delta Y_t - E(\Delta Y_t|D_{t-1})$ on $\Delta D_t - E(\Delta D_t|D_{t-1})$, and an FE for the second FD. The conditional expectations in each period are estimated using Lasso. The coefficient on $\Delta D_t$ is equal to $-0.54$, it is significant (s.e.=0.14), and statistically different from the paper's coefficient ($H=7.18$). Using NN to estimate the conditional expectations, we obtain a coefficient equal to $-0.50$, different from zero (s.e.=0.15), and statistically different from the paper's coefficient ($H=12.24$).

\paragraph{Re-analysis of the reduced-form of the IV regression in the paper's Table 2 Column (7).}
On top of FD OLS regressions, \cite{acemoglu2016import} also estimate FD IV regressions, where $\Delta D$ is instrumented by $\Delta Z$, the change in industries' Chinese-import-penetration ratio in eight other high-income countries. We reanalyze the reduced-form regression attached to the two-period FD IV in Table 2 Column (7) of the paper, namely a regression of $\Delta Y$ on $\Delta Z$ taking first differences from 1999 to 2007 as before. The coefficient on $\Delta Z$ is negative and significant ($-1.40$, s.e.=0.38).  Instead, we regress $\Delta Y - E(\Delta Y|Z_1)$ on $\Delta Z - E(\Delta Z|Z_1)$, where the conditional expectations are estimated using Lasso. The coefficient is smaller than before and significant ($-0.86$, s.e.=0.35), and significantly different from the paper's reduced-form coefficient ($H = 5.15$). When we reproduce the exercise using NN, the slope coefficient equals $-0.67$,  it is significantly different from zero (s.e.=0.35) and from the paper's reduced-form coefficient ($H = 6.78$). If we control for $E(\Delta Z|Z_1)$ linearly, the coefficient on $\Delta Z$ is not significantly different from zero.

\begin{sidewaystable}
\begin{table}[H]
\centering
 \caption{First-difference regressions of employment on Chinese imports}\label{table:AADH}
\begin{tabular}{l l  l l l c c c}
\hline
Time periods & Dep var & Indep var & Learner & Estimated in paper?  & Coefficient & s.e. & N
                                             \\\hline
1999 to 2007 & $\Delta Y$  & $\Delta D$ & None & No & $-0.78$  & $0.22$ & 392 \\
1999 to 2007 & $\Delta Y$  & $\Delta D$, $D_1$ & None & No & $-0.65$  & $0.24$ & 392 \\
 1999 to 2007 & $\Delta Y-E(\Delta Y|D_1)$ & $\Delta D-E(\Delta D|D_1)$ &  Lasso & No & $-0.60$  & $0.21$ & 392 \\
 1999 to 2007 & $\Delta Y-E(\Delta Y|D_1)$ & $\Delta D-E(\Delta D|D_1)$ & NN & No & $-0.65$  & $0.21$ & 392 \\
 1999 to 2007 &  $\Delta Y_{-1}$ & $\Delta D$ & None & No & $-0.23$  & $0.15$ & 392 \\
  1999 to 2007 &  $\Delta Y_{-1}$ & $\Delta D$, $D_0$, $D_1$ & None & No & $-0.11$  & $0.14$ & 392 \\
  1999 to 2007 &  $\Delta Y_{-1}-E(\Delta Y_{-1}|D_0,D_1)$& $\Delta D-E(\Delta D|D_0,D_1)$ & Lasso & No & $-0.04$  & $0.15$ & 392 \\
    1999 to 2007 &  $\Delta Y_{-1}-E(\Delta Y_{-1}|D_0,D_1)$& $\Delta D-E(\Delta D|D_0,D_1)$ & NN & No & $-0.08$  & $0.13$ & 392 \\
Stacked & $\Delta Y_t$ & $\Delta D_t$ & None & Table 2 Column (2) & $-0.81$  & $0.16$ & 784 \\
Stacked & $\Delta Y_t$ & $\Delta D_t$, $D_{t-1}$ & None & No & $-0.60$  & $0.15$ & 784 \\
Stacked & $\Delta Y_t-E(\Delta Y_t|D_{t-1})$  & $\Delta D_t-E(\Delta D_t|D_{t-1})$ & Lasso & No & $-0.54$  & $0.14$ & 784 \\
Stacked & $\Delta Y_t-E(\Delta Y_t|D_{t-1})$  & $\Delta D_t-E(\Delta D_t|D_{t-1})$ & NN & No & $-0.50$  & $0.15$ & 784 \\
 1999 to 2007 &  $\Delta Y$ & $\Delta Z$ & None & RF of Table 2 Column (7) & $-1.40$  & $0.38$ & 392 \\
 1999 to 2007 & $\Delta Y$ & $\Delta Z, Z_{1}$ & None & No & $-0.55$  & $0.35$ & 392 \\
  1999 to 2007 & $\Delta Y-E(\Delta Y|Z_1)$ & $\Delta Z-E(\Delta Z|Z_1)$ & Lasso & No & $-0.86$  & $0.35$ & 392 \\
    1999 to 2007 &  $\Delta Y-E(\Delta Y|Z_1)$ & $\Delta Z-E(\Delta Z|Z_1)$ & NN & No & $-0.68$  & $0.35$ & 392 \\
\hline
\end{tabular}
\begin{minipage}{16.0cm}
\footnotesize{The table shows regressions of US industries employment evolutions $\Delta Y$ on the evolution of their Chinese import-penetration ratio $\Delta D$, using the data of \cite{acemoglu2016import}. Regressions are weighted by industries' 1991
employment, and standard errors are clustered on 135 three-digit industries, as in the aforementioned paper. The first line shows a FD regression from 1999 to 2007. The second, third, and fourth lines are similar, but control for $E(\Delta D|D_1)$. The second simply assumes a linear model. The third and fourth use a locally-robust estimator, where conditional expectations are estimated by Lasso and NN, respectively. The fifth line performs a naive placebo test. The sixth, seventh, and eighth lines conduct placebo tests controlling for $E(\Delta D|D_0,D_1)$, where conditional expectations are estimated by a linear model, Lasso and NN, respectively.  The ninth line re-estimates the stacked regression in the paper's Table 2 Column (2). The tenth, eleventh, and twelfth lines are similar to the ninth, but control for $E(\Delta D_t|D_{t-1})$, either linearly, or using a locally-robust estimator, where conditional expectations are estimated by Lasso and NN, respectively. The thirteenth line re-estimates the reduced-form of the paper's IV regression in Table 2 Column (7). The last three lines are similar, but control for $E(\Delta Y|Z_1)$  linearly, or using a locally-robust estimator, where conditional expectations are estimated by Lasso and NN, respectively.}
\end{minipage}
\end{table}
\end{sidewaystable}

\section{Conclusion}

When the baseline treatment varies, first-difference regressions may suffer from an omitted variable bias if the treatment effect changes over time. A simple fix is to control for $E(\Delta D|D_1)$.

\newpage

\bibliography{biblio}

\begin{thebibliography}{}

\bibitem[\protect\citeauthoryear{Acemoglu, Autor, Dorn, Hanson, and
  Price}{Acemoglu et~al.}{2016}]{acemoglu2016import}
Acemoglu, D., D.~Autor, D.~Dorn, G.~H. Hanson, and B.~Price (2016).
\newblock Import competition and the great us employment sag of the 2000s.
\newblock {\em Journal of Labor Economics\/}~{\em 34\/}(S1), S141--S198.

\bibitem[\protect\citeauthoryear{Algan and Cahuc}{Algan and
  Cahuc}{2010}]{algan2010}
Algan, Y. and P.~Cahuc (2010).
\newblock Inherited trust and growth.
\newblock {\em American Economic Review\/}~{\em 100\/}(5), 2060--2092.

\bibitem[\protect\citeauthoryear{Angrist}{Angrist}{1998}]{angrist1998}
Angrist, J.~D. (1998).
\newblock Estimating the labor market impact of voluntary military service
  using social security data on military applicants.
\newblock {\em Econometrica\/}~{\em 66\/}(2), 249--288.

\bibitem[\protect\citeauthoryear{Angrist and Pischke}{Angrist and
  Pischke}{2009}]{angrist2009mostly}
Angrist, J.~D. and J.-S. Pischke (2009).
\newblock {\em Mostly harmless econometrics: An empiricist's companion}.
\newblock Princeton university press.

\bibitem[\protect\citeauthoryear{Arkhangelsky, Imbens, Lei, and
  Luo}{Arkhangelsky et~al.}{2021}]{arkhangelsky2021double}
Arkhangelsky, D., G.~W. Imbens, L.~Lei, and X.~Luo (2021).
\newblock Double-robust two-way-fixed-effects regression for panel data.
\newblock {\em arXiv preprint arXiv:2107.13737\/}.

\bibitem[\protect\citeauthoryear{Athey and Imbens}{Athey and
  Imbens}{2022a}]{athey2021design}
Athey, S. and G.~W. Imbens (2022a).
\newblock Design-based analysis in difference-in-differences settings with
  staggered adoption.
\newblock {\em Journal of Econometrics\/}~{\em 226}, 62--79.

\bibitem[\protect\citeauthoryear{Athey and Imbens}{Athey and
  Imbens}{2022b}]{athey2022design}
Athey, S. and G.~W. Imbens (2022b).
\newblock Design-based analysis in difference-in-differences settings with
  staggered adoption.
\newblock {\em Journal of Econometrics\/}~{\em 226\/}(1), 62--79.

\bibitem[\protect\citeauthoryear{Autor, Dorn, and Hanson}{Autor
  et~al.}{2013}]{autor2013china}
Autor, D.~H., D.~Dorn, and G.~H. Hanson (2013).
\newblock The china syndrome: Local labor market effects of import competition
  in the united states.
\newblock {\em American economic review\/}~{\em 103\/}(6), 2121--2168.

\bibitem[\protect\citeauthoryear{Blandhol, Bonney, Mogstad, and
  Torgovitsky}{Blandhol et~al.}{2022}]{blandhol2022tsls}
Blandhol, C., J.~Bonney, M.~Mogstad, and A.~Torgovitsky (2022).
\newblock When is tsls actually late?
\newblock NBER working paper 29709.

\bibitem[\protect\citeauthoryear{Borusyak and Hull}{Borusyak and
  Hull}{2023}]{borusyak2023quasi}
Borusyak, K. and P.~Hull (2023).
\newblock On quasi-experimental shift-share iv with heterogeneous treatment
  effects.
\newblock Technical report, Technical report.

\bibitem[\protect\citeauthoryear{Borusyak and Hull}{Borusyak and
  Hull}{2024}]{borusyak2024negative}
Borusyak, K. and P.~Hull (2024).
\newblock Negative weights are no concern in design-based specifications.
\newblock In {\em AEA Papers and Proceedings}, Volume 114, pp.\  597--600.
  American Economic Association 2014 Broadway, Suite 305, Nashville, TN 37203.

\bibitem[\protect\citeauthoryear{Borusyak, Hull, and Jaravel}{Borusyak
  et~al.}{2022}]{borusyak2022quasi}
Borusyak, K., P.~Hull, and X.~Jaravel (2022).
\newblock Quasi-experimental shift-share research designs.
\newblock {\em The Review of economic studies\/}~{\em 89\/}(1), 181--213.

\bibitem[\protect\citeauthoryear{Chernozhukov, Chetverikov, Demirer, Duflo,
  Hansen, Newey, and Robins}{Chernozhukov
  et~al.}{2018}]{chernozhukov2018double}
Chernozhukov, V., D.~Chetverikov, M.~Demirer, E.~Duflo, C.~Hansen, W.~Newey,
  and J.~Robins (2018).
\newblock Double/debiased machine learning for treatment and structural
  parameters.

\bibitem[\protect\citeauthoryear{Chernozhukov, Escanciano, Ichimura, Newey, and
  Robins}{Chernozhukov et~al.}{2022}]{chernozhukov2022locally}
Chernozhukov, V., J.~C. Escanciano, H.~Ichimura, W.~K. Newey, and J.~M. Robins
  (2022).
\newblock Locally robust semiparametric estimation.
\newblock {\em Econometrica\/}~{\em 90\/}(4), 1501--1535.

\bibitem[\protect\citeauthoryear{Chor, Manova, and Yu}{Chor
  et~al.}{2021}]{chor2021growing}
Chor, D., K.~Manova, and Z.~Yu (2021).
\newblock Growing like china: Firm performance and global production line
  position.
\newblock {\em Journal of International Economics\/}~{\em 130}, 103445.

\bibitem[\protect\citeauthoryear{de~Chaisemartin and
  D'Haultf{\oe}uille}{de~Chaisemartin and D'Haultf{\oe}uille}{2020}]{dcDH2020}
de~Chaisemartin, C. and X.~D'Haultf{\oe}uille (2020).
\newblock Two-way fixed effects estimators with heterogeneous treatment
  effects.
\newblock {\em American Economic Review\/}~{\em 110\/}(9), 2964--2996.

\bibitem[\protect\citeauthoryear{de~Chaisemartin and
  D'Haultf{\oe}uille}{de~Chaisemartin and
  D'Haultf{\oe}uille}{2021}]{de2020difference}
de~Chaisemartin, C. and X.~D'Haultf{\oe}uille (2021).
\newblock Difference-in-differences estimators of intertemporal treatment
  effects.
\newblock arXiv preprint arXiv:2007.04267.

\bibitem[\protect\citeauthoryear{de~Chaisemartin, D'Haultfoeuille, Pasquier,
  and Vazquez-Bare}{de~Chaisemartin et~al.}{2022}]{chaisemartin2022continuous}
de~Chaisemartin, C., X.~D'Haultfoeuille, F.~Pasquier, and G.~Vazquez-Bare
  (2022).
\newblock Difference-in-differences for continuous treatments and instruments
  with stayers.
\newblock arXiv preprint arXiv:2201.06898.

\bibitem[\protect\citeauthoryear{de~Chaisemartin, D'Haultf{\oe}uille, and
  Vazquez-Bare}{de~Chaisemartin et~al.}{2024}]{de2024difference}
de~Chaisemartin, C., X.~D'Haultf{\oe}uille, and G.~Vazquez-Bare (2024).
\newblock Difference-in-difference estimators with continuous treatments and no
  stayers.
\newblock In {\em AEA Papers and Proceedings}, Volume 114, pp.\  610--613.
  American Economic Association 2014 Broadway, Suite 305, Nashville, TN 37203.

\bibitem[\protect\citeauthoryear{de~Chaisemartin and Lei}{de~Chaisemartin and
  Lei}{2021}]{de2021more}
de~Chaisemartin, C. and Z.~Lei (2021).
\newblock More robust estimators for instrumental-variable panel designs, with
  an application to the effect of imports from china on us employment.
\newblock {\em arXiv preprint arXiv:2103.06437\/}.

\bibitem[\protect\citeauthoryear{Duranton and Turner}{Duranton and
  Turner}{2011}]{Duranton2011}
Duranton, G. and M.~A. Turner (2011, October).
\newblock The fundamental law of road congestion: Evidence from us cities.
\newblock {\em American Economic Review\/}~{\em 101\/}(6), 2616--2652.

\bibitem[\protect\citeauthoryear{Gentzkow, Shapiro, and Sinkinson}{Gentzkow
  et~al.}{2011}]{gentzkow2011}
Gentzkow, M., J.~M. Shapiro, and M.~Sinkinson (2011).
\newblock The effect of newspaper entry and exit on electoral politics.
\newblock {\em American Economic Review\/}~{\em 101\/}(7), 2980--3018.

\bibitem[\protect\citeauthoryear{Imbens and Xu}{Imbens and
  Xu}{2024}]{imbens2024lalonde}
Imbens, G. and Y.~Xu (2024).
\newblock Lalonde (1986) after nearly four decades: Lessons learned.
\newblock arXiv preprint arXiv:2406.00827.

\bibitem[\protect\citeauthoryear{Imbens, Rubin, and Sacerdote}{Imbens
  et~al.}{2001}]{imbens2001estimating}
Imbens, G.~W., D.~B. Rubin, and B.~I. Sacerdote (2001).
\newblock Estimating the effect of unearned income on labor earnings, savings,
  and consumption: Evidence from a survey of lottery players.
\newblock {\em American economic review\/}~{\em 91\/}(4), 778--794.

\bibitem[\protect\citeauthoryear{Ishimaru}{Ishimaru}{2022}]{ishimaru2022we}
Ishimaru, S. (2022).
\newblock What do we get from a two-way fixed effects estimator? implications
  from a general numerical equivalence.
\newblock {\em Preprint, submitted October\/}~{\em 18}.

\bibitem[\protect\citeauthoryear{Robins}{Robins}{1986}]{robins1986new}
Robins, J. (1986).
\newblock A new approach to causal inference in mortality studies with a
  sustained exposure period-application to control of the healthy worker
  survivor effect.
\newblock {\em Mathematical modelling\/}~{\em 7\/}(9-12), 1393--1512.

\bibitem[\protect\citeauthoryear{Robinson}{Robinson}{1988}]{robinson1988root}
Robinson, P.~M. (1988).
\newblock Root-n-consistent semiparametric regression.
\newblock {\em Econometrica: Journal of the Econometric Society\/}, 931--954.

\bibitem[\protect\citeauthoryear{Wooldridge}{Wooldridge}{2010}]{wooldridge2010econometric}
Wooldridge, J.~M. (2010).
\newblock {\em Econometric analysis of cross section and panel data}.
\newblock MIT press.

\bibitem[\protect\citeauthoryear{Yitzhaki}{Yitzhaki}{1996}]{yitzhaki1996}
Yitzhaki, S. (1996).
\newblock On using linear regressions in welfare economics.
\newblock {\em Journal of Business \& Economic Statistics\/}~{\em 14\/}(4),
  478--486.

\end{thebibliography}

\newpage

\section{Supplemental Appendix, Not For Publication}

\subsection{If the treatment effect is non linear, $\beta_{D_1}$ identifies a weighted average of derivatives of $x\mapsto Y_2(x)$.}

Without a linear-treatment-effect assumption, we replace Assumption \ref{hyp:cond_random_fd} by the following assumption:
\begin{hyp}\label{hyp:cond_random_fd_nl}
\begin{equation}
\Delta D \indep (Y_2(0)-Y_1(0),(Y_1(d_1)-Y_1(0))_{d_1\in \mathcal{D}_1},(Y_2(x)-Y_2(0))_{x\in \mathcal{D}_2}) |D_1,
\end{equation}
\end{hyp}
where $\mathcal{D}_1$ and $\mathcal{D}_2$ respectively denote the supports of $D_1$ and $D_2$. Without a linear treatment effect assumption, rather than just assuming that $\Delta D$ is independent of the slopes $S_1$ and $S_2$ conditional on $D_1$, we need to assume that $\Delta D$ is independent of $Y_1(d_1)-Y_1(0)$, the effect of increasing the treatment from $0$ to $d_1$ on the period-one outcome, and of $Y_2(x)-Y_2(0)$, the effect of increasing the treatment from $0$ to $x$ on the period-two outcome, for all $d_1$ and $x$.
 \begin{thm}\label{thm:new_nl}
Suppose that the distribution of $D_2|D_1=d_1$ is continuous for all $d_1\in \mathcal{D}_1$, that $x\mapsto E(\Delta Y|D_1=d_1,D_2=x)$ is differentiable wrt $x$ for all $d_1\in \mathcal{D}_1$ and all $x$ in the support of $D_2|D_1=d_1$, that $x\mapsto Y_2(x)$ is differentiable wrt $x$ for all $x$ in the support of $D_2$ with $|\frac{\partial Y_2(x)}{\partial x}|\leq M$ for some real number $M$,  and that Assumption \ref{hyp:cond_random_fd_nl} holds. Then, $$\beta_{D_1}=\frac{E\left(\int_{\mathcal{D}_2}w_{x,D_1}E\left(\frac{\partial Y_2(x)}{\partial x}\middle|D_1=d_1\right)dx\right)}{E\left(\int_{\mathcal{D}_2}w_{x,D_1}dx\right)},$$
where $w_{x,D_1}=(E(D_2|D_1,D_2\geq x)-E(D_2|D_1,D_2<x))P(D_2\geq x|D_1)(1-P(D_2\geq x|D_1)).$
\end{thm}
\textbf{Proof}\\
As $\beta_{D_1}=E(\Delta Y(D_2-E(D_2|D_1)))/E(D_2(D_2-E(D_2|D_1)))$, it follows from Equation (3.3.10) in \cite{angrist2009mostly}\footnote{That result is a generalization of the result in \cite{yitzhaki1996} to a regression with control variables.} that
$$\beta_{D_1}=\frac{E\left(\int_{\mathcal{D}_2}w_{x,D_1}\frac{\partial E(\Delta Y|D_1,D_2=x)}{\partial x}dx\right)}{E\left(\int_{\mathcal{D}_2}w_{x,D_1}dx\right)}.$$
Then, for all $d_1\in \mathcal{D}_1$ and all $x$ in the support of $D_2|D_1=d_1$,
\begin{align*}
&E(\Delta Y|D_1=d_1,D_2=x)\\
=&E(Y_2(x)-Y_1(d_1)|D_1=d_1,D_2=x)\\
=&E(Y_2(0)-Y_1(0)|D_1=d_1,D_2=x)+E(Y_2(x)-Y_2(0)|D_1=d_1,D_2=x)\\
-&E(Y_1(d_1)-Y_1(0)|D_1=d_1,D_2=x)\\
=&E(Y_2(0)-Y_1(0)|D_1=d_1)+E(Y_2(x)-Y_2(0)|D_1=d_1)-E(Y_1(d_1)-Y_1(0)|D_1=d_1),
\end{align*}
where the third equality follows from Assumption \ref{hyp:cond_random_fd_nl}.
Therefore,
\begin{align*}
\frac{\partial E(\Delta Y|D_1,D_2=x)}{\partial x}=&\frac{\partial E(Y_2(x)|D_1=d_1)}{\partial x}\\
=& E\left(\frac{\partial Y_2(x)}{\partial x}\middle|D_1=d_1\right),
\end{align*}
where the second equality follows from the dominated convergence theorem. The result follows by plugging the previous display into the displayed formula for $\beta_{D_1}$ above \textbf{QED.}

\subsection{Estimation of $\beta_{D_1,Y_1}$ in the empirical application}

	\begin{table}[H]
		\centering
		\caption{Regressions of $Y_2$ on $D_2$ and $E\left(D_2 \mid D_1, Y_1\right)$}\label{table:AADH2}
		\begin{tabular}{l l  l l l c c c}
			\hline
			Time periods & Dep var & Indep var & Learner  & Coefficient & s.e. & N
			\\\hline
			2007 & $Y_2$   & $D_2$, $D_1$, $Y_1$ & None & $-0.62$  & $0.24$ & 392 \\
			2007 & $Y_2$   & $D_2 - E\left(D_2 \mid D_1, Y_1\right)$  & Lasso & $-0.59$  & $0.20$ & 392 \\
				2007 & $Y_2$   & $D_2 - E\left(D_2 \mid D_1, Y_1\right)$ & NN & $-0.78$  & $0.29$ & 392 \\
			\hline
		\end{tabular}
		\begin{minipage}{16.0cm}
			\footnotesize{The table shows regressions of US industries employment in 2007, $Y_2$, on Chinese import-penetration ratio in 2007, $D_2$, using the data of \cite{acemoglu2016import}. Regressions are weighted by industries' 1991
				employment, and standard errors are clustered on 135 three-digit industries, as in the aforementioned paper. All these specifications control for the augmented regression $E(D_2|D_1, Y_1)$. The first line assumes that $E(D_2|D_1, Y_1)$ is linear. The second and third use a locally-robust estimator, where conditional expectations are estimated by Lasso and NN, respectively.}

	\end{minipage}
\end{table}

\end{document}